\documentclass[aps,pre,reprint,groupedaddress]{revtex4-2}

\usepackage{graphicx}
\usepackage{times}
\usepackage{amssymb}
\usepackage{amsmath}
\usepackage[table]{xcolor}
\usepackage{hyperref}

\newcommand{\nn}{\nonumber}
\newcommand{\vphi}{\varphi}
\newcommand{\veps}{\varepsilon}
\newcommand{\ea}{eqnarray}

\begin{document}

\title{Electroprewetting near a flat charged surface}

\author{Yoav Tsori}
\email[]{tsori@bgu.ac.il}
\homepage[]{www.bgu.ac.il/~tsori}
\affiliation{Department of Chemical Engineering, Ben-Gurion University of the 
Negev, Israel.}


\begin{abstract}

We look at the wetting of a pure fluid in contact with a charged flat surface. 
In the bulk, the fluid is a classical van der Waals fluid containing  
dissociated ions. The presence of wall and ions leads to strong 
dielectrophoretic and electrophoretic forces that increase the fluid's density 
at the wall. We calculate the fluid's profiles analytically and numerically 
and obtain the energy integrals. The critical surface potential for prewetting 
is obtained. In the phase diagrams, the line of first-order transition meets a 
second-order transition line at a critical point whose temperature can be higher 
or lower than the bulk critical temperature. The results are relevant to 
droplet nucleation around charged particles in the atmosphere and could possibly 
explain deviations from expected nucleation rates.

\end{abstract}

\maketitle

\section{Introduction}

Fluids near or at charged surfaces are ubiquitous in every day life and in 
technology. The depth of penetration of the field into the fluid can vary from 
the molecular scale to infinity, in principle, depending on the fluids 
conductivity and on the geometry of electrodes producing the field. In the first 
edition of their book on electrodynamics of continuous media, Landau and 
Lifshitz calculated in a typical concise manner the change to the critical 
temperature $T_c$ \cite{ll_book_elec}. Their reasoning was as follows: The 
electrostatic energy of perfect dielectrics is given by the integral $F_{\rm 
e}=-(1/2)\int\veps_0\veps(\phi){\bf E}^2{\rm d}^3r$, where ${\bf E}$ is the 
electric field, $\veps_0$ is the vacuum permittivity, and $\veps(\phi)$ is the 
relative dielectric constant, that depends on the scaled density (or 
composition, in the analogous case of binary mixtures), $\phi$. Close enough to 
the critical point, one can write $\veps(\phi)$ as a Taylor-series expansion to 
quadratic order around $\phi_c$: 
$\veps(\phi)\simeq\veps_c+\Delta\veps(\phi-\phi_c)+(1/2)\veps''(\phi-\phi_c)^2$, 
where $\phi_c$ is the value of $\phi$ at the critical point, and 
$\veps_c=\veps(\phi_c)$. When $\veps(\phi)$ is multiplied by ${\bf E}^2$ in the 
electrostatic energy integral, and ${\bf E}$ is constant, the linear term in 
$\phi-\phi_c$ is simply a renormalization of the chemical potential. 
In a Landau expansion of the total energy in powers of 
$\phi-\phi_c$, the quadratic term in $F_{\rm e}$, proportional to $\veps''$, 
simply renormalizes $T_c$, such that $T_c\to T_c+\Delta T$. For binary 
liquid mixtures,  
one obtains $\Delta T=v_0\veps_0\veps''E^2/k_B$, where $v_0$ is a molecular 
volume and $k_B$ is Boltzmann's constant. Consequently, $T_c$ and the whole 
binodal curve are shifted upwards or downwards depending on the sign of 
$\veps''$. For a pure liquid in coexistence with its vapor the behavior is 
similar. For values of $v_0$, $\veps''$ and $E$ characteristic to experiments, 
one finds that $\Delta T$ is quite small, $\Delta T\sim 10$ mK. 

The Landau prediction spurred several very accurate 
measurements of the shift to $T_c$, starting from P. Debye  
\cite{debye_jcp1965,reich_jpspp1979,early_jcp1992,wirtz_fuller_prl1993,
orzech_chemphys1999,tsori_jcpb_2014}. 
But due to the smallness of the effect and some errors in 
several works, the problem gradually lost its allure. 

Spatially nonuniform fields, on the other hand, have a much stronger effect on 
the phase behavior of fluids \cite{tsori_rmp2009,tsori_pnas2007}. When a 
spatially nonuniform field acts on a liquid mixture, if the field gradient is 
small, then the dielectrophoretic force ``pulls'' the more polar component 
toward the region where the field is high. If the external potential is small, 
then this effect is ``obvious'' in the sense that composition gradients are 
smooth. However, when the external potential is large enough, a sharp interface 
appears between two coexisting domains, one rich in the polar component and the 
other rich in the less polar liquid. The interface can appear far from a wall or 
at a surface, and the phase transition can be first or second order, depending 
on $T$, average composition $\phi_0$, and symmetry 
\cite{tsori_pre2013,tsori_jcp2014a,tsori_jcp_2009,tsori_jpcb_2011}. The shift of 
the stability line separating the homogeneous and mixed states in field 
gradients is $10$--$100$ larger than in uniform fields (Landau). 

In purely dielectric liquids (e.g., oils), the dielectrophoretic force is 
governed by the geometry of the electrodes producing the fields. In polar 
liquids on the other hand, dissociated ions are free to move. Their location is 
a balance among entropic, electrostatic, and chemical forces and it generally 
leads to screening of the field. Thus, field gradients occur irrespective of 
the electrode geometry. Importantly, when the ions move to the electrode they 
also ``drag'' the liquid in which they are better solvated, leading to a force 
of electrophoretic origin which is proportional to the preferential solvation. 

The Gibbs transfer energy $\Delta G^t$ measures this preference and is  
sensitive to the ion and solvent combination. For example, for Na$^+$ cations, 
$\Delta G^t=8.6$ kJ/mol for moving from water to methanol (Na$^+$ ``prefers'' to 
be in water) but $\Delta G^t=-13.7$ kJ/mol for moving from water to dimethyl 
sulfoxide (it ``prefers'' to be in DMSO). The corresponding values for the 
hydrophilic Cl$^-$ anion are both positive ($13.1$ and $39.4$ kJ/mol, 
respectively), and the values for the bulky hydrophobic anion sodium 
tetraphenylborate are $-24$ kJ/mol and $-37$ kJ/mol, respectively 
\cite{hefter_pac2005,onuki_cocis2011b}. Ion-specific effects play a central role 
in the solubility of proteins and leads to ``salting out'' \cite{zhang2006}, 
which is commonly used in protein separation techniques. Ionic specificity leads 
to the Hofmeister series 
\cite{levin_langmuir2010,sivan_langmuir2013,netz_langmuir2013, 
netz_langmuir2014} and influences the air-water surface tension and the 
interaction between surface 
\cite{jungwirth2006,levin_prl2009a,levin_prl2009b,andelman_jcp2015, 
andelman_jcp2011}, \cite{marcus_book}.

In this paper, we study the coexistence of a vapor and 
liquid near a flat charged surface. The origin of field gradients is the 
presence of ions dissociated in the fluid. We look in details on the 
electroprewetting phenomenon, obtain the critical surface potential (or charge) 
for the phase transition, calculate the profiles, and show the resulting phase 
diagrams.

\section{Model}

The starting point of the field-theoretic model is the free-energy integral 
\begin{eqnarray}
F&=&\int f {\rm d}{\bf r}+\int_S f_{\rm s}{\rm d}{\bf r}_s,\nn
\end{eqnarray}
where $f$ and $f_s$ are the bulk free energy and surface energy densities:
\begin{eqnarray}\label{FE}
f&=&f_{\rm vdw}(T,\rho)+\frac12 
c^2(\nabla\rho)^2+k_BT\left[n^+(\ln(v_0n^+)-1)\right.\nn\\
&+&\left. n^-(\ln(v_0n^-)-1)\right]
-\frac{1}{2}\veps_0\veps(\rho)(\nabla\psi)^2\nn\\
&+&e(n^+-n^-)\psi-k_BTv_0(\Delta u^+n^++\Delta u^-n^-)\rho,\nn\\
f_{\rm s}&=&v_0\rho\gamma_{\rm ls}+(1-v_0\rho)\gamma_{\rm 
gs}.
\end{eqnarray}
Here $\rho$ is the coarse-grained fluid's density, $n^\pm$ are the cation and 
anion densities, $\veps_0$ is the vacuum permittivity and $\veps(\rho)$ is the 
relative dielectric constant, $\psi$ is the electrostatic potential, $k_B$ is 
the Boltzmann's constant, $T$ is the absolute temperature, $v_0$ is a 
molecular volume (assumed the same for all molecules involved), $e$ is the unit 
charge, and $c$ is a constant. The surface integral is over the bounding 
surfaces; in this work we consider a semi-infinite system in the $x>0$ plane 
bounded by a single surface at $x=0$. The parameters $\gamma_{\rm ls}$ and 
$\gamma_{\rm gs}$ are the liquid-solid and gas-solid interfacial energies, 
respectively. The parameters $\Delta u^\pm$ measure the preferential solubility 
of the ions in the liquid. When they are positive, ions are attracted to the 
liquid phase, where $\rho$ is high. The derivation below is general for a 
bistable fluid, but for numerical purposes we will consider specifically the 
classical van der Waals fluid where $f_{\rm vdw}$ is given by
\begin{eqnarray}
f_{\rm vdw}=k_BT\rho\left[\ln(\rho\Lambda^3)-1-\ln(1-\rho 
b)\right]-a\rho^2.
\end{eqnarray}
Here $a$ and $b$ are the van der Waals interaction and excluded volume 
parameters, and $\Lambda$ is the de Broglie wavelength. The bulk critical 
point of the classical van der Waals fluid is given by $T_c=8a/27k_B b$, 
$\rho_c=1/3b$, and $P_c=a/27b^2$. 

We proceed by defining dimensionless quantities as follows: $\tilde{f}=f/P_c$, 
$\phi=\rho/\rho_c$, $t=T/T_c$, and $\tilde{n}^\pm=v_0n^\pm$, 
$\tilde{\psi}=e\psi/k_BT$. In addition, all lengths are scaled as $\tilde{{\bf 
r}}={\bf r}/\lambda_{D0}$, where  $\lambda_{D0}$ is the ``vacuum'' Debye length 
given by
\begin{\ea}
\lambda_{D0}^2=\frac{\veps_0v_0k_BT}{2\tilde{n}_0e^2}
\end{\ea}
and $\tilde{n}_0=v_0n_0$ is the scaled bulk ion concentration, far away from 
any charged surface. Lastly, we define $\tilde{P}$, $M$, ``vacuum'' Bjerrum 
length $l_B$ and $\tilde{c}$ as
\begin{\ea}
\tilde{P}&=&\frac{k_BT}{v_0P_c},~~~~~~
M=\tilde{P}\frac{v_0}{l_B\lambda_{D0}^2},\nn\\
l_B&=&\frac{e^2}{\veps_0k_BT},~~~~~~
\tilde{c}^2=\frac{c^2\rho_c^2}{P_c\lambda_{D0}^2}.
\end{\ea}
The dimensionless bulk energy integral is therefore
\begin{\ea}\label{dimensionless_bulk_FE}
\frac{F_b}{\lambda_{D0}^3P_c}&=&\int\left(\tilde{f}_{\rm 
vdw}(\phi)+\frac12\tilde{c}^2(\tilde{\nabla}\phi)^2\right.\nn\\
&+&\tilde{P}[\tilde{n}^+(\ln(\tilde{n}^+)-1)
+\tilde{n}^-(\ln(\tilde{n}^-)-1)]\nn\\
&-&\frac12 M\veps(\phi)(\tilde{\nabla} \tilde{\psi})^2
+\tilde{P}(\tilde{n}^+-\tilde{n}^-)\tilde{\psi}\nn\\
&-&\left.\tilde{P}v_0\rho_c(\Delta u^+\tilde{n}^++\Delta u^-\tilde{n}^-)\phi
\right){\rm d}\tilde{{\bf r}},
\end{\ea}
with
\begin{\ea}
\tilde{f}_{\rm vdw}=(8t/3)\phi\left[\ln(\rho_c\phi
\Lambda ^ 3 ) - 1 -\ln(1-\phi/3)\right]-3\phi^2.
\end{\ea}
For the surface energy, we repeat similar steps and obtain
\begin{\ea}\label{dimensionless_surf_FE}
\frac{F_s}{\lambda_{D0}^3P_c}&=&\int 
\Delta\tilde{\gamma}\phi d\tilde{{\bf r}}_s+{\rm const}\\
\Delta\tilde{\gamma}&=&\frac{v_0\rho_c(\gamma_{\rm ls}-\gamma_{\rm 
gs})}{\lambda_{D0}P_c}.\nn
\end{\ea}
%

In equilibrium, the total free energy $F_b+F_s$ is at a minimum, and thus we 
look at the Euler-Lagrange equations for 
$\tilde{\omega}=\tilde{f}-\mu\phi-\mu^+\tilde{n}^+-\mu^-\tilde{n}^-$ with 
respect to the four fields $\phi$, $\tilde{n}^\pm$, and $\psi$:
\begin{\ea}
\frac{\delta \tilde{\omega}}{\delta \phi}&=&
\tilde{f}'_{\rm vdw}(\phi)-\tilde{c}^2\tilde{\nabla}^2\phi
-\frac12 M\frac{d\veps}{d\phi}(\tilde{\nabla}\tilde{\psi})^2\nn\\
&-&\tilde{P}v_0\rho_c(\Delta u^+\tilde{n}^++\Delta 
u^-\tilde{n}^-)-\mu=0,\label{eq_el1}\\
\frac{\delta \tilde{\omega}}{\delta 
n^\pm}&=&\tilde{P}[\ln(\tilde{n}^\pm)\pm\tilde{\psi}-v_0\rho_c\Delta 
u^\pm\phi]-\mu^\pm=0,\label{eq_el2}\\
\frac{\delta \tilde{\omega}}{\delta 
\psi}&=&M\tilde{\nabla}[\veps(\phi)\tilde{\nabla}\tilde{\psi}
]+\tilde{P}(\tilde{n}^+-\tilde{n}^-)=0.\label{eq_el3}
\end{\ea}
$\mu$, $\mu^+$ and $\mu^-$ are the chemical potentials of the fluid density and 
the cation and anion densities, respectively. Equation (\ref{eq_el3}) is the 
Poisson's equation.

For a fluids in contact with a wall at scaled potential $\tilde{V}$ and bulk 
composition $\phi_0$, the boundary conditions are
\begin{\ea}
\tilde{c}^2\phi'(0)&=&\Delta\tilde{\gamma},~~~\phi(\infty)=\phi_0\nn\\
\tilde{\psi}(0)&=&\tilde{V},~~~~~\tilde{\psi}(\infty)=0.\nn
\end{\ea}
In the grand-canonical ensemble, the chemical potential is $\mu=\mu_0$ with
\begin{\ea}
\mu_0&=&\tilde{f}'_{\rm vdw}(\phi_0)-2\tilde{P}\tilde{n}_0v_0\rho_c\Delta u.
\end{\ea}

We continue by focusing on the simple case where the two ions are equally 
``philic'' to the liquid phase: $\Delta u^\pm=\Delta u$. 
We use $\veps_1\approx 1$ and $\veps_2$ as the permittivities of the vapor 
and pure liquid, respectively, and assume a dielectric constitutive relation of 
the form
\begin{\ea}
\veps(\phi)=\veps_1+\Delta\tilde{\veps}\phi+\frac12\veps''\phi^2.
\end{\ea}
To show a clear difference from the Landau mechanism, in most places 
below we will assume a linear relation with $\veps''=0$ and 
$\Delta\veps=(\veps_2-\veps_1)/3$.

Far away from any charged surface, in the bulk reservoir, the electric potential 
vanishes, $\phi=\phi_0$ is a composition smaller than the lower binodal 
value, and $\tilde{n}=\tilde{n}_0$, hence the Euler-Lagrange equation 
for the ions is readily solved to yield the Boltzmann weight
\begin{\ea}
\tilde{n}^\pm=\tilde{n}_0e^{\mp\tilde{\psi}+v_0\rho_c\Delta u(\phi-\phi_0)}.
\end{\ea}

\section{Results}

\subsection{Critical value of the potential}

The dielectrophoretic and electrophoretic forces in Eq. (\ref{eq_el1}) lead to 
an increase of the fluid's density near the wall. This increase is reminiscent 
of the increase in density of a fluid under the influence of gravity near the 
surface of Earth (though the electric field case is far richer because the 
fluid's density changes the field, via the Poisson's equation, a ``back 
action'' that is not present in the coupling between gravity and mass 
density) \cite{moldover_rmp_1979}. 

Because the fluid is bistable, if the surface potential is large enough, 
then a transition will occur from a gas to a liquid phase. To obtain the
threshold value we substitute 
$\tilde{n}^++\tilde{n}^-=2\tilde{n}_0e^{v_0\rho_c\Delta 
u(\phi-\phi_0)}\cosh(\tilde{\psi})$ in Eq. (\ref{eq_el1}) and obtain
\begin{\ea}
\tilde{f}'_{\rm vdw}(\phi)&-&\tilde{c}^2\tilde{\nabla}^2\phi
-\frac12 M\Delta\veps(\tilde{\nabla}\tilde{\psi})^2\\
&-&2\tilde{P}\tilde{n}_0v_0\rho_c\Delta ue^{v_0\rho_c\Delta 
u(\phi-\phi_0)}\cosh(\tilde{\psi})-\mu_0=0.\nn\label{eq_el4}
\end{\ea}
When $\veps_1=\veps_2$, the solution to 
the Poisson-Boltzmann problem [Eqs. (\ref{eq_el2})--(\ref{eq_el3})] is the 
classical nonlinear potential near a single wall: 
\begin{\ea}\label{psi0}
\tilde{\psi}_0&=&2\ln\left[\frac{1+Ce^{-\tilde{x}}}{1-Ce^{-\tilde{x}}}\right],
~~~~~~C=\frac{e^{\tilde{V}/2}-1}{e^{\tilde{V}/2}+1}.
\end{\ea}
When $\veps_1\neq\veps_2$ and $\Delta\veps$ is much smaller than the 
average dielectric constant $\bar{\veps}$, the solution to the potential is 
$\tilde{\psi}=\tilde{\psi}_0+\tilde{\psi}_1$, where $\tilde{\psi}_1$ is 
of order $\Delta\veps/\bar{\veps}$. The field squared 
$(\tilde{\nabla}\tilde{\psi})^2$ in Eq.~(\ref{eq_el4}) can be expanded 
similarly in orders of $\Delta\veps/\bar{\veps}$. It is instructive to examine 
this equation to zero order in $\Delta\veps/\bar{\veps}$ and in the absence of 
preferential solvation ($\Delta u=0$); corrections will be the result of 
a loop expansion in higher orders in $\Delta\veps/\bar{\veps}$.

\begin{figure}[!th]
\begin{center}
\includegraphics[width=0.47\textwidth,bb=90 170 525 605,clip]{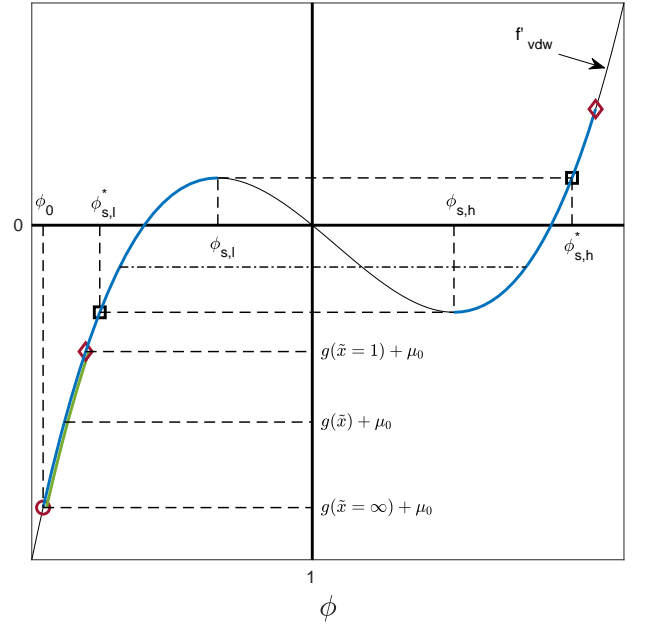}
\caption{Graphical solution of Eq.~(\ref{eq_el1}) or Eq. (\ref{eq_el4}) when
$\veps_1\approx\veps_2$ and $\Delta u=0$. These equations can be written as 
$f'_{\rm vdw}=g(\tilde{x})+\mu_0$, where 
$g(\tilde{x})=(1/2)M\Delta\veps(\tilde{\nabla}\tilde{\psi})^2$ is a decreasing 
function of $\tilde{x}$. Curved black line is $f'_{\rm vdw}$. The solution to 
the equation is the intersection between $f'_{\rm vdw}$ and a nearly horizontal 
line. This line sweeps upwards as $\tilde{x}$ decreases from infinity, giving a 
smoothly decaying profile $\phi(\tilde{x})$ (green segment) when the surface 
potential $\tilde{V}$ is small. Red circle is the bulk composition and leftmost 
red diamond is the surface composition $\phi_s$. If $\tilde{V}$ is large enough, 
then coexistence between a vapor and a liquid is marked by a surface 
composition at large values of $\phi$ (rightmost red diamond).
}
\label{fig1}
\end{center}
\end{figure}

Cahn's classical wetting construction cannot be performed here due to the 
existence of the long-range force \cite{cahn_jcp_1977,pgg_rmp_1985}, but 
graphical solution of the equation for $\phi$ may be useful in the sharp 
interface limit ($\tilde{c}=0$), see Fig. \ref{fig1}. For simplicity, it is 
assumed here that $\veps$ is linear in $\phi$ and that $\Delta u=0$. The black 
thin line is $\tilde{f}'_{\rm vdw}(\phi)$ at a given temperature $t<1$. $\phi_0$ 
is the bulk composition and is assumed to be outside and on the left of the 
binodal curve. The solution to Eq.~(\ref{eq_el4}) is given graphically as an 
intersection between this curve and a nearly horizontal line 
$(1/2)M\Delta\veps(\tilde{\nabla}\tilde{\psi})^2+\mu_0$. At $\tilde{x}=\infty$, 
the field is zero and $\tilde{f}'_{\rm vdw}=\mu_0$ retrieves the bulk phase (red 
circle in the figure). As $\tilde{x}$ decreases, the field increases, the 
horizontal line sweeps upwards in Fig. \ref{fig1}, and the intersection moves to 
larger values of $\phi$. If the wall charge is not too large, then the locus of 
intersections is shown as a green solid line. The red diamond signifies the 
maximal value of $\phi$, obtained at the wall: $\phi_s=\phi(\tilde{x}=0)$. The 
resulting profile $\phi(\tilde{x})$ is smoothly decaying with $\tilde{x}$.

This behavior changes at large wall potentials. Let us denote by $\phi_{\rm 
s,l}$ and $\phi_{\rm s,h}$ the low and high spinodal compositions, respectively. 
$\phi_{\rm s,l}^*$ is the composition where $f'_{\rm vdw}$ equals $f'_{\rm 
vdw}(\phi_{\rm s,h})$. Similarly, $\phi_{\rm s,h}^*$ is the composition where 
$f'_{\rm vdw}(\phi_{\rm s,h}^*)=f'_{\rm vdw}(\phi_{\rm s,l})$. Clearly, when the 
potential or charge increases sufficiently, the line 
$(1/2)M\Delta\veps(\tilde{\nabla}\tilde{\psi})^2+\mu_0$ moves up so much that 
the solution to the equation is at large values of $\phi$. At these 
potentials, the profile is described by a ``jump'' from a low-$\phi$ part at 
large $\tilde{x}$ to a high-$\phi$ part at small $\tilde{x}$ . The compositions 
in both phases are not uniform. At the critical value for the transition, 
$\phi_s$ is larger than $\phi_{\rm s,l}^*$ and smaller than $\phi_{\rm s,l}$.

One can approximate the critical potential by assuming that the transition 
occurs at the lower binodal value, that is, when $\phi_s=\phi_{\rm b,l}$.
Since $\left(4C/(C^2-1)\right)^2=2(\cosh(\tilde{V})-1)$, and in the sharp 
interface limit, this amounts to 
\begin{\ea}
&&\tilde{f}'_{\rm vdw}(\phi_{\rm b,l})-M\Delta\veps(\cosh(\tilde{V}_c)-1)\nn\\
&&-2\tilde{P}\tilde{n}_0v_0\rho_c\Delta ue^{v_0\rho_c\Delta 
u(\phi_{\rm b,l}-\phi_0)}\cosh(\tilde{V}_c)-\mu_0\approx 0,\nn
\end{\ea}
or
\begin{\ea}\label{eq_Vc_sa}
\cosh(\tilde{V}_c)\approx \frac{\tilde{f}'_{\rm 
vdw}(\phi_{\rm 
b,l})-\mu_0+M\Delta\veps}{M\Delta\veps+2\tilde{P}\tilde{n}
_0v_0\rho_c\Delta ue^{v_0\rho_c\Delta u(\phi_{\rm b,l}-\phi_0)}}.\nn\\
\end{\ea}
When $\tilde{V}\gg 1$, $\cosh(\tilde{V})-1\approx \cosh(\tilde{V})$ and 
we have 
\begin{\ea}\label{eq_Vc_sa2}
\cosh(\tilde{V}_c)\approx\frac{\tilde{f}'_{\rm 
vdw}(\phi_{\rm b,l})-\mu_0}{M\Delta\veps+2\tilde{P}\tilde{n}
_0v_0\rho_c\Delta ue^{v_0\rho_c\Delta u(\phi_{\rm b,l}-\phi_0)}}.\nn\\
\end{\ea}
In the opposite limit, $\tilde{V}_c\ll 1$, one obtains
\begin{\ea}\label{eq_Vc_sa3}
\tilde{V}_c^2\approx\frac{\tilde{f}'_{\rm 
vdw}(\phi_{\rm b,l})-2\tilde{P}\tilde{n}_0v_0\rho_c\Delta ue^{v_0\rho_c\Delta 
u(\phi_{\rm 
b,l}-\phi_0)}-\mu_0}{\frac12 M\Delta\veps+\tilde{P}\tilde{n}
_0v_0\rho_c\Delta ue^{v_0\rho_c\Delta u(\phi_{\rm b,l}-\phi_0)}}.\nn\\
\end{\ea}
\begin{figure}[!th]
\begin{center}
\includegraphics[width=0.47\textwidth,bb=80 225 490 550,clip]{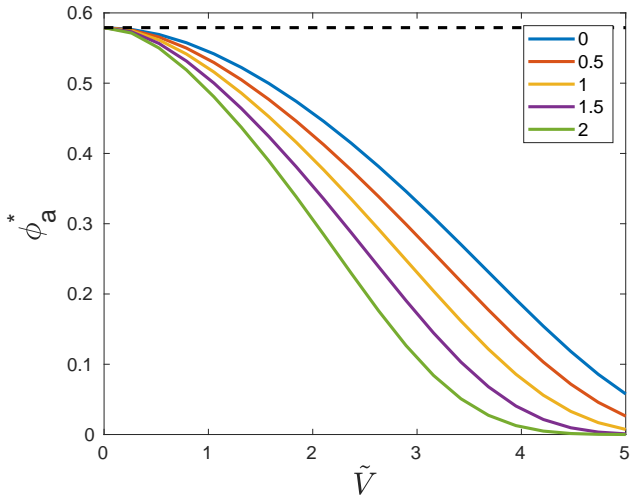}
\caption{Plots of $\phi_a^*$, the value of the bulk composition $\phi_0$ 
obeying Eq. (\ref{eq_Vc_sa}), as a function of surface potential $\tilde{V}$ 
and for a fixed reduced temperature $t=T/T_c=0.95$. The five curves correspond 
to the five values of $v_0\rho_c\Delta u$ indicated in the legend. Wetting of 
the surface occurs when $\phi_0>\phi_a^*$. Dashed horizontal line is the lower 
value of the binodal composition $\phi_{b,l}$ at the same temperature. The 
$\tilde{V}\ll 1$ and $\tilde{V}\gg 1$ limits are described by Eqs. 
(\ref{eq_Vc_sa3}) and (\ref{eq_Vc_sa2}), respectively. In this and in other 
figures (unless stated otherwise) we used $\tilde{n}_0=0.05$, $\veps_1=1$, 
$\veps_2=5$, and $\tilde{P}/t=0.4$.
}
\label{fig2}
\end{center}
\end{figure}
The analytical equation Eq. (\ref{eq_Vc_sa}) can be inverted to obtain the 
critical value of bulk composition $\phi_a^*$ for a given temperature and 
surface potential. Figure \ref{fig2} shows $\phi_a^*$ vs $\tilde{V}$. When 
$\phi_0<\phi_a^*$, the profile $\phi(\tilde{x})$ is smoothly varying;
when $\phi_0>\phi_a^*$, an electroprewetting transition occurs where a dense 
phase wets the surface in coexistence with a vapor phase away from it. 
At infinitesimal potential, Eq. (\ref{eq_Vc_sa3}) applies and $\phi_a^*$ 
is infinitesimally close to the 
lower binodal value $\phi_{b,l}$. As $\tilde{V}$ increases, $\phi_a^*$ 
decreases and the unstable ``window'' for electroprewetting 
$\phi_{b,l}-\phi_a^*$, widens. At large potentials Eq. (\ref{eq_Vc_sa2}) 
holds.

\subsection{Phase diagrams}

\begin{figure}[!th]
\begin{center}
\includegraphics[width=0.47\textwidth,bb=35 100 550 680,clip]{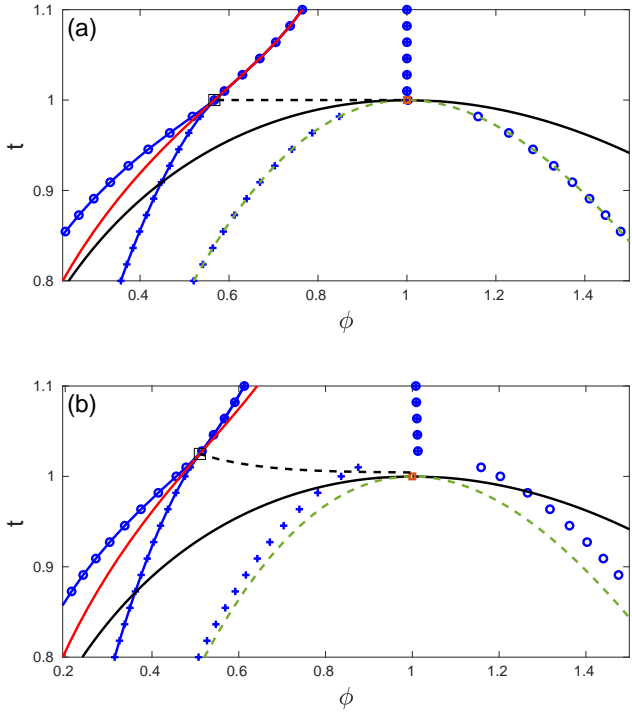}
\caption{A solution of Eqs. 
(\ref{surf_eq1}) and (\ref{surf_eq2}) for the variables $\phi_0$ (bulk 
composition) and $\phi_s$ (surface composition) gives lines in the phase diagram 
(blue). These equations have two pairs of solutions -- one is marked with 
circles [circles with line: $\phi_0(t)$; circles: $\phi_s(t)$] and one 
marked with ``+'' signs [``+'' with line:  $\phi_0(t)$; ``+'':
$\phi_s(t)$]. In (a) $v_0\rho_c\Delta u=0$ and $\phi_s(t)$ are identical with 
the bulk spinodal curves (dashed green line). In (b) $v_0\rho_c\Delta u=0.8$ and 
they are different. Red curve $\phi_0(t)$ is given from from Eqs. 
(\ref{surf_eq1}) and (\ref{surf_eq3}). This curve meets with the blue curves at 
a point marked with a square. In (a) this occurs at $t=1$ while in (b) it occurs 
at $t>1$. In both parts, the point marked with square is the simultaneous 
solution of Eqs. (\ref{surf_eq1})--(\ref{surf_eq3}) for a given surface 
potential $\tilde{V}$ [in both (a) and (b) $\tilde{V}=2$]. The dashed line, 
obtained as the solution for $\tilde{V}$ diminishing to zero, is the 
approximation for the second-order phase transition line. The bulk critical 
point of the van der Waals fluid is at $(1,1)$. In Figs. \ref{fig3}--\ref{fig9} 
we used $v_0\rho_c\Delta u=0.2$.
}
\label{fig3}
\end{center}
\end{figure}

Before we evaluate the phase diagrams numerically let us continue with the 
sharp interface limit and examine the Euler-Lagrange equation for 
$\phi$ along with the second and third derivatives of the energy calculated 
{\it at the surface}:
\begin{\ea}
f'_{\rm vdw}(\phi_s)-\frac12 
M\frac{d\veps}{d\phi}\psi_s'^2\nn\\
-2\tilde{P}\tilde{n}_0v_0\rho_c\Delta 
ue^{v_0\rho_c\Delta 
u(\phi_s-\phi_0)}\cosh(\tilde{V})-\mu_0&=&0~\label{surf_eq1},\\
f''_{\rm vdw}(\phi_s)-\frac12 
M\frac{d^2\veps}{d\phi^2}\psi_s'^2\nn\\
-2\tilde{P}\tilde{n}_0(v_0\rho_c\Delta 
u)^2e^{v_0\rho_c\Delta u(\phi_s-\phi_0)}\cosh(\tilde{V})&=&0~\label{surf_eq2},\\
f'''_{\rm vdw}(\phi_s)-\frac12 
M\frac{d^3\veps}{d\phi^3}\psi_s'^2\nn\\-2\tilde{P}\tilde{n}_0(v_0\rho_c\Delta 
u)^3e^{v_0\rho_c\Delta u(\phi_s-\phi_0)}\cosh(\tilde{V})&=&0~\label{surf_eq3}.
\end{\ea}
where $-\psi_s'$ is the value of the field at $\tilde{x}=0$. A simultaneous 
solution of Eqs. (\ref{surf_eq1}) and (\ref{surf_eq2}) yields $\phi_0(t)$ and 
$\phi_s(t)$ as two lines in the phase diagram. Simultaneous solution of Eqs. 
(\ref{surf_eq1}) and (\ref{surf_eq3}) yields another two lines $\phi_0(t)$ and 
$\phi_s(t)$. Figure \ref{fig3} shows these lines for the case where $\veps$ is a 
linear function of $\phi$ and with the approximation 
$\tilde{\psi}=\tilde{\psi}_0$ from Eq.~(\ref{psi0}). There are two solutions to 
Eqs. (\ref{surf_eq1}) and (\ref{surf_eq2}). The first is marked with circles 
(circles with line: $\phi_0$, circles only: $\phi_s$) and the second is marked 
with a plus ``+'' sign (``+'' with line: $\phi_0$; ``+'' only: $\phi_s$). Black 
solid line is the bulk binodal curve. In part (a) the preferential solvation is 
zero ($\Delta u=0$) and $\phi_s$ are identical with the spinodal composition 
(dashed green line). Red solid curve is the solution for $\phi_0$ of Eqs. 
(\ref{surf_eq1}) and (\ref{surf_eq3}). 

The curves $\phi_0(t)$ for the second and third derivatives merge at a 
critical point marked with square. At this point the line of first order 
prewetting transitions becomes a line of second order transitions 
\cite{saam_prl_1987}. When $\Delta u=0$, the temperature of this point $t=1$ is 
equal to the bulk critical temperature. Above this temperature, the lines merge 
into one line. The point marked with square is thus a simultaneous solution of 
the three equations (\ref{surf_eq1})--(\ref{surf_eq3}) for the three variables 
$\phi_0$, $\phi_s$, and $t$. The dashed line leading from this point to the bulk 
critical point is the second-order transition line appearing in the presence of 
long-range forces. It is approximately obtained as the locus of square points 
with surface potentials decreasing from $\tilde{V}$ to zero. In part (b), we 
show the same curves but now $\Delta u\neq 0$. The curves $\phi_s$ are different 
from the bulk spinodal values. The point where first- and second-order 
transitions meet is shifted upwards to $t>1$.

In the above analysis, we used the approximation 
$\tilde{\psi}=\tilde{\psi}_0$ which is valid when $\veps_1$ and $\veps_2$ are 
sufficiently similar. In general, when the two epsilons are very different
energy penalties can be associated with dielectric interfaces that are 
perpendicular to the field's direction. In our problem, it means that the 
effect of $\tilde{\psi}_1$, which is of order $\Delta\veps/\bar{\veps}$ and 
neglected in the previous discussion, becomes increasingly important. The 
approximation is worse if, in addition, $\veps$ depends quadratically on $\phi$. 
In the $\tilde{\psi}=\tilde{\psi}_0$ approximation, the point where first- and 
second-order lines meet is predicted too far to the left of the binodal curve 
(small values of $\phi_0$) in the phase diagram.
\begin{figure}[!th]
\begin{center}
\includegraphics[width=0.47\textwidth,bb=30 80 550 705,clip]{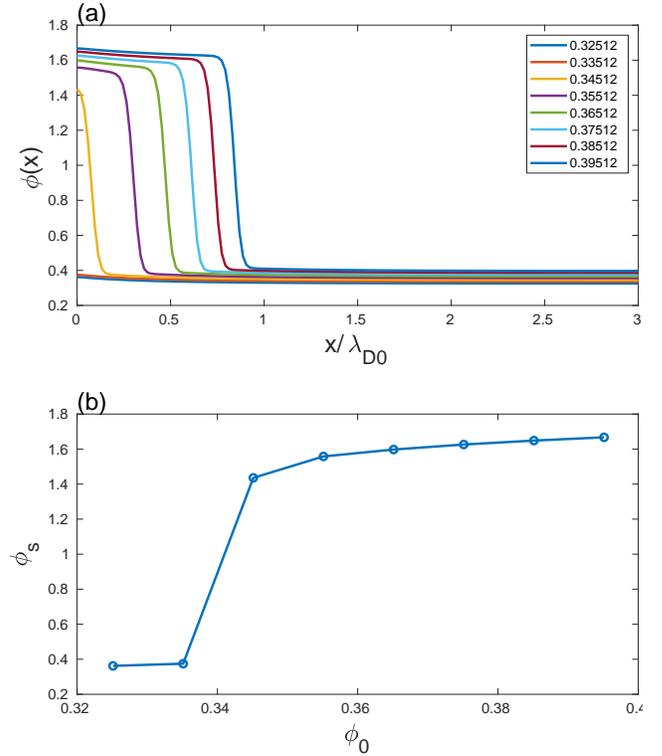}
\caption{(a) Composition profiles $\phi(\tilde{x})$ for a fluid confined by 
a charged wall at $\tilde{x}=0$ for several values of bulk composition $\phi_0$ 
(see legend), at fixed temperature and surface voltage. Below the critical 
composition $\approx 0.34$ the surface is wetted by a vapor while above this 
composition it is wetted by a liquid. The liquid-vapor interface is always at a 
finite distance but increases as the composition approaches the binodal value. 
(b) Surface composition $\phi_s$ vs the bulk composition $\phi_0$ in part (a). 
We used $\tilde{V}=2$, $\tilde{c}=0.02$, $\Delta\tilde{\gamma}=0$, and 
$t=0.9$.
}
\label{fig4}
\end{center}
\end{figure}

We now lift this assumption and use full profiles calculated numerically. 
Figure \ref{fig4} (a) depicts $\phi(\tilde{x})$ for several bulk compositions at 
fixed $t$ and $\tilde{V}$. At the two lower compositions, the profiles are 
smoothly varying and the surface has vapor phase. Above the critical composition 
(in this example $\phi_0\approx 0.34$), a first-order transition occurs and the 
surface is wet by the liquid. At even higher compositions, the vapor-liquid 
interface moves to larger (but always finite) distances.
Part (b) shows the surface composition $\phi_s$ vs $\phi_0$.
\begin{figure}[!th]
\begin{center}
\includegraphics[width=0.47\textwidth,bb=40 130 550 635,clip]{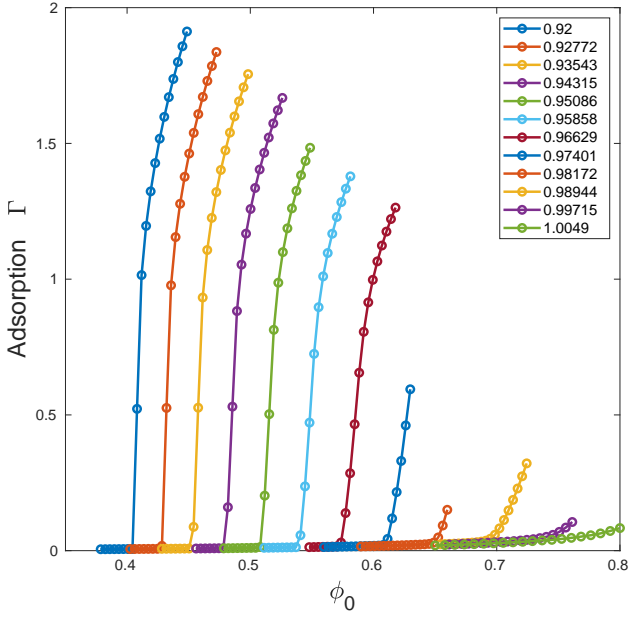}
\caption{Adsorption curves $\Gamma(\phi_0)$ at different temperatures $t$ given 
in the legend. The abrupt change in $\Gamma(\phi_0)$ is the 
first-order prewetting transition. We used $\tilde{V}=1$, $\tilde{c}=0.02$, 
and $\Delta\tilde{\gamma}=0$.
}
\label{fig5}
\end{center}
\end{figure}

Figure \ref{fig5} shows the adsorption curves $\Gamma$ at different 
temperatures and fixed potential. $\Gamma$ is defined as 
\begin{\ea}
\Gamma=\int_0^\infty \phi(\tilde{x})d\tilde{x}.
\end{\ea}
For all temperatures, the values of $\phi_0$ used are smaller than the 
corresponding binodal composition. $\Gamma$ increases with $\phi_0$ 
monotonically and has an abrupt change at a specific composition, 
signaling the prewetting transition. At temperatures closer to the bulk 
critical point ($t\lesssim 1$) the change in $\Gamma$ is less abrupt. 
Across the second-order line the change is continuous but this is not 
discernible at the parameters used here and is not shown.

\begin{figure}[!th]
\begin{center}
\includegraphics[width=0.47\textwidth,bb=30 180 550 590,clip]{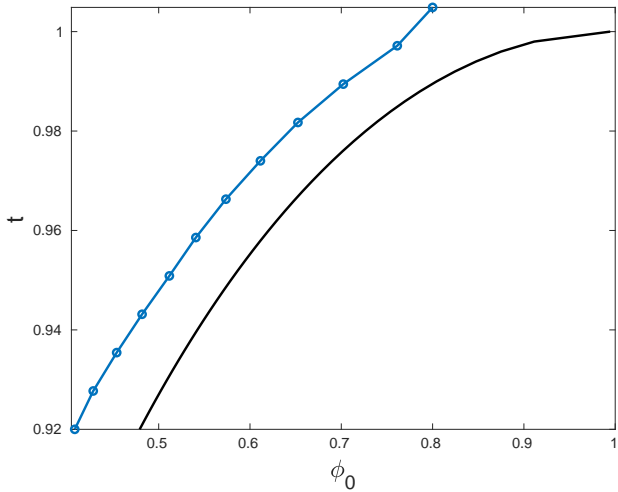}
\caption{Phase diagram at constant potential. Black line is the binodal 
curve and blue line is the electro-prewetting line. It terminates 
at a temperature that is close to $1$ but slightly larger because  
$\Delta u\neq 0$. We used $\tilde{V}=1$, $\tilde{c}=0.02$, $v_0\rho_c\Delta 
u=0.2$, and $\Delta\tilde{\gamma}=0$.
}
\label{fig6}
\end{center}
\end{figure}

Figure \ref{fig6} is the phase diagram in the $(\phi_0,t)$ plane for a 
fixed value of $\tilde{V}$ and a diffuse interface ($\tilde{c}\neq 0$). 
In these axes, the critical point of the classic van der Waals fluid is at 
$(1,1)$. The black concave curve is the bulk 
field-free binodal obtained from the common-tangent construction.
The blue line is the first-order transition. 
Compositions to its left lead to a dry surface while to its right the surface 
becomes wet. The first order line terminates at a temperature where the 
transition becomes second order (see Fig. \ref{fig3}). In the numerical scheme 
using a diffuse interface, $\tilde{c}\neq 0$, it is difficult to reliably find 
this line and hence it is not shown.

\subsection{Analytical profiles and quadratic expansion of the 
energy}

We now seek analytical approximations for the composition profiles and the film 
energy. At surface potentials smaller than the critical value, the profiles 
$\phi(\tilde{{\bf r}})$ are smooth and the variation $|\phi(\tilde{{\bf 
r}})-\phi_0|$ is everywhere small. In that case, the free-energy Eq. 
(\ref{dimensionless_bulk_FE}) can be expanded to quadratic order in both 
$\vphi\equiv \phi-\phi_0$ and $\tilde{\psi}^2$. The total grand-canonical 
potential $\Omega$ is 
\begin{\ea}\label{fe_expansion}
\frac{\Omega}{\lambda_{D0}^3P_c}&\approx & \phi_0\Delta\tilde{\gamma} +
\int\left\{\tilde{f}_{\rm 
vdw}(\phi_0)-2\tilde{P}\tilde{n}_0-\mu_0\phi_0\right.\nn\\
&+&\frac12 
\alpha^2\tilde{c}^2\vphi^2+\frac12\tilde{c}^2(\tilde{\nabla}\vphi)^2 -\frac12 
M\left(\veps(\phi_0)+\Delta \veps\vphi\right)(\tilde{\nabla} 
\tilde{\psi})^2\nn\\
&-&\left.\tilde{P}\tilde{n}_0[1+v_0\rho_c\Delta 
u\vphi]\tilde{\psi}^2
\right\}{\rm d}\tilde{{\bf r}}+\int\vphi(\tilde{{\bf 
r}}_s)\Delta\tilde{\gamma}{\rm d}\tilde{{\bf r}}_s,\nn\\
\end{\ea}
where 
\begin{\ea}
\alpha^2&=&\frac{\tilde{f}_{\rm 
vdw}''(\phi_0)-2\tilde{P}\tilde{n}_0(v_0\rho_c\Delta u)^2}{\tilde{c}^2}~
\end{\ea}
is the inverse squared correlation length modified by the presence of the ions 
(see Appendix A for details). The quantity $\alpha^2$ replace the standard 
bulk $f''(\phi)$ term in a Landau-series expansion; it contains a correction to 
the correlation length due to the preferential solubility of the ions and hints 
to a change in the bulk critical point \cite{onuki_prl2017}.

The first line in Eq. (\ref{fe_expansion}) includes constant terms; the last 
term in the third line is the change in the surface energy, which is linear in 
$\vphi$. 
For the specific case of a van der Waals fluid, the vanishing of 
the second derivative is written as 
\begin{\ea}
\frac{8t}{3}\left(\frac{1}{\phi}+\frac{1}{3-\phi}\right)+\frac{8t}{(3-\phi)^
2}-6-2\tilde{P}\tilde{n}_0(v_0\rho_c\Delta u)^2=0.\nn
\end{\ea}
In the bulk, the field vanishes and the third derivative of the free energy is 
\begin{\ea}
\frac{72t\left(1-\phi\right)}{\left(3-\phi\right)^3\phi^2}
-2\tilde{P}\tilde{n}_0(v_0\rho_c\Delta u)^3=0.\nn
\end{\ea}
Treating $v_0\rho_c\Delta u$ as small and looking to linear order in $\Delta t$ 
and $\Delta\phi_c$, we find that the bulk critical point is shifted 
according to 
\begin{\ea}\label{t_shift_pref_sol}
t_c&=&1+\Delta t~~~~~~~~~~~~~~~~~~~~~~~~~~~\phi_c=1+\Delta\phi_c,\\
\Delta t&=&\frac13\tilde{P}\tilde{n}_0(v_0\rho_c\Delta 
u)^2~~~~~~~~~\Delta\phi_c=\frac{2}{9}\tilde{P}\tilde{n}_0(v_0\rho_c\Delta 
u)^3.\nn
\end{\ea}
One can estimate the magnitude of this shift for water by using 
$P_c=22\times 10^{6}$ Pa, $T_c=670^\circ$K, $v_0\simeq3\times 10^{-29}$ m$^3$, 
$\rho_c\simeq 10^{28}$ m$^{-3}$, and taking $\Delta u=1$ 
and $\tilde{n}_0=0.001$, that is, one ion on every $1000$ water molecules. It 
then follows that $\Delta T=(1/3)T_c\tilde{P}\tilde{n}_0(v_0\rho_c\Delta 
u)^2\simeq 0.1^\circ$K.
\begin{figure}[!th]
\begin{center}
\includegraphics[width=0.47\textwidth,bb=55 120 530 650,clip]{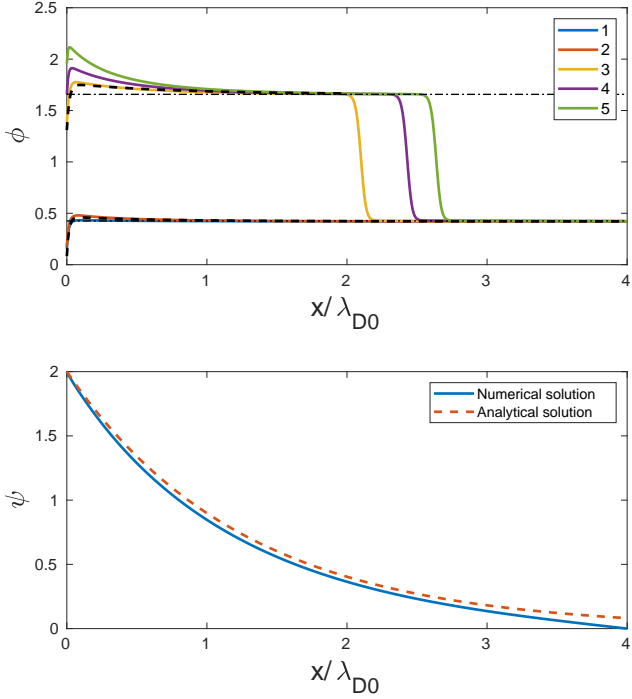}
\caption{(a) Composition profiles at several values of the surface potential 
$\tilde{V}$ (see legend). For the two small values of $\tilde{V}$, the profile 
$\phi(\tilde{x})$ smoothly decreases with increasing $\tilde{x}$. At short 
distance,  $\phi(\tilde{x})$ increases since $\Delta\tilde{\gamma}=0.01$ is 
positive, favoring the vapor phase. When the surface potential is large enough, 
$\tilde{V}=3$, $4$, $5$, an electroprewetting transition occurs, with liquid 
phase at small $\tilde{x}$ coexisting with a vapor at larger distances. Fine 
dash-dotted horizontal lines are the low and high values of the binodal 
compositions at the same temperature, $\phi_{b,l}$ and $\phi_{b,h}$, 
respectively. Lower and higher dashed lines are the linear approximations based 
on  Eq. (\ref{lin_approx}). (b) The numerical (solid) and analytical (dashed) 
potential profiles $\tilde{\psi}(\tilde{x})$ for $\tilde{V}=2$ of part (a). In 
both parts, the reduced temperature is $t=0.9$ and $\tilde{c}=0.02$.
}
\label{fig7}
\end{center}
\end{figure}

We now take the variation of the quadratic energy with respect to $\vphi$ and 
$\tilde{\psi}$. To lowest order we obtain
\begin{\ea}
&&\alpha^2\tilde{c}^2\vphi-\tilde{c}^2\vphi''-\frac12 
M\Delta\tilde{\veps}(\tilde{\nabla}\tilde{\psi})^2
-\tilde{P}\tilde{n}_0v_0\rho_c\Delta u\tilde{\psi}^2=0,\nn\\~\\
&&M\veps(\phi_0)\tilde{\nabla}^2\tilde{\psi}
-2\tilde{P}\tilde{n}_0\tilde{\psi}=0.
\end{\ea}
In one dimension, the Poisson equation is readily solved noting that 
$2\tilde{P}\tilde{n}_0/M=1$:
$\tilde{\psi}=\tilde{V}e^{-\beta\tilde{x}}$, where $\beta^2=1/\veps(\phi_0)$.
The equation for $\vphi$ can now be written as
\begin{\ea}
\vphi''-\alpha^2\vphi&=&Be^{-2\beta\tilde{x}},
\end{\ea}
where
\begin{\ea}
B&=&-\frac{M}{2\tilde{c}^2}
\left(v_0\rho_c\Delta u+\Delta\veps\beta^2 \right)\tilde{V}^2.\nn
\end{\ea}
The form of the quantity $B$ shows that the preferential solvation, 
proportional to $\Delta u$, leading to an electrophoretic force on the fluid, 
appears on the same footing as the ``dielectric contrast'', 
$\Delta\veps\beta^2$, in our notation, which leads to a 
dielectrophoretic force. 
\begin{figure}[!th]
\begin{center}
\includegraphics[width=0.47\textwidth,bb=30 190 555 600,clip]{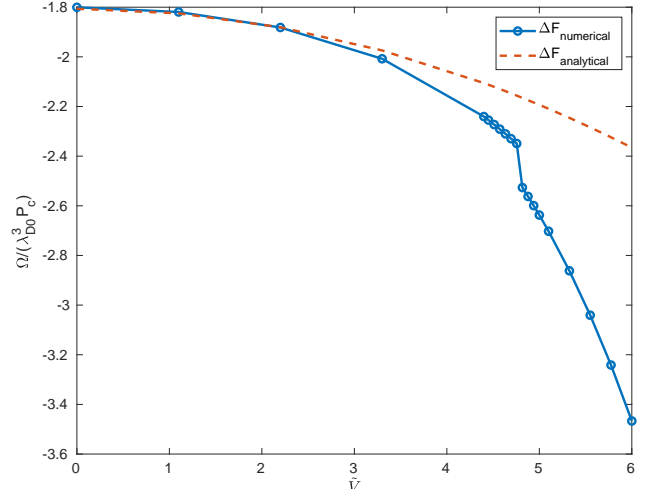}
\caption{Numerical [solid line, from Eqs. (\ref{dimensionless_bulk_FE}) and 
(\ref{dimensionless_surf_FE})] and analytical [dashed line, Eq. 
(\ref{quadratic_approx})] film energies as a function of increasing surface 
potential $\tilde{V}$. The energy decreases monotonically as $\tilde{V}$ 
increases. For the values of parameters chosen here, the critical 
electroprewetting potential is $\tilde{V}_c\approx 4.75$. When 
$\tilde{V}<\tilde{V}_c$, the analytical approximation matches well the numerical 
energy; above $\tilde{V}_c$ it is not a good approximation as the film energy 
exhibits a sharp decrease in its value. We used $\Delta\tilde{\gamma}=-0.005$, 
$t=0.7$, $\tilde{c}=0.1$, and $\phi_0=0.99\phi_{b,l}$ for this temperature.
}
\label{fig8}
\end{center}
\end{figure}

The solution for $\vphi$ obeying the boundary conditions 
$\tilde{c}^2\vphi'(0)=\Delta\tilde{\gamma}$ and 
$\vphi(\infty)=0$ is
\begin{\ea}\label{lin_approx}
\vphi&=&Ae^{-\alpha\tilde{x}}+\frac{B}{4\beta^2-\alpha^2}e^{-2\beta\tilde{x}},
\end{\ea}
with
\begin{\ea}\label{A}
A&=&-\frac{2\beta 
B}{\alpha(4\beta^2-\alpha^2)}-\frac{\Delta\tilde{\gamma}}{\alpha\tilde{c}^2}.
\end{\ea}
Note that the deviation of the surface composition from the bulk value is 
\begin{\ea}
\vphi(0)=-\frac{\Delta\tilde{\gamma}}{\alpha\tilde{c}^2}-\frac{B}{\alpha(
\alpha+2\beta)}.
\end{\ea}
Since $B$ is negative, when $\Delta\tilde{\gamma}=0$ it follows that 
$\vphi(0)>0$: The dielectrophoretic and electrophoretic forces 
are both attracting the liquid to the wall increasing its density there.

We are now in a position to substitute the profiles just obtained back in the 
energy, to get the integrated energy (see Appendix B for details):
\begin{\ea}\label{en_integ}
\frac{\Omega_{\rm 
tot}-\Omega_0}{\lambda_{D0}^3P_c}&=&-\frac12M\tilde{V}^2/\beta
+\frac12\tilde{c}^2A^2\alpha-2\tilde{c}^2\frac{AB
\beta}{\alpha^2-4\beta^2}\nn\\
&-&\frac18\tilde{c}^2\frac{B^2}{\beta}\frac{\alpha^2
-12\beta^2}{\left(\alpha^2-4\beta^2\right)^2}
+\left(A+\frac{B}{4\beta^2-\alpha^2}\right)\Delta\tilde{\gamma}.\nn\\
\end{\ea}
$\Omega_0/(\lambda_{D0}^3P_c)=\phi_0\Delta\tilde{\gamma}+\int_0^\infty[
\tilde{f}_{\rm vdw}(\phi_0)-2\tilde{P}\tilde{n}_0-\mu_0\phi_0]{\rm 
d}\tilde{x}$ is the (infinite) bulk energy.

We further use the relation between $A$ and $B$ in Eq. (\ref{A}) to 
eliminate $A$ and express $\Omega_{\rm tot}$ in terms of $B$ and 
$\Delta\tilde{\gamma}$:
\begin{\ea}\label{quadratic_approx}
\frac{\Omega_{\rm 
tot}-\Omega_0}{\lambda_{D0}^3P_c}&=&
-\frac12M\tilde{V}^2/\beta-\frac18\frac{B^2\tilde{c}^2}{\alpha\beta}\frac{
\alpha+4\beta}{(\alpha+2\beta)^2}\nn\\
&-&\frac{B\Delta\tilde{\gamma } }
{\alpha(\alpha+2\beta)}-\frac12\frac{\Delta\tilde{\gamma}^2}{\alpha\tilde{c}^2}.
\end{\ea}

In the absence of surface potential, $\tilde{V}=0$ and $B\propto\tilde{V}^2=0$, 
and the energy is just the last term. This negative energy is the standard 
expression, quadratic in $\Delta\tilde{\gamma}$. When the surface has no 
chemical wetting preference, $\Delta\tilde{\gamma}=0$, the energy is quadratic 
in the potential. In the general case, a mixed term $\propto 
B\Delta\tilde{\gamma}$ exists. 

Figure \ref{fig7} shows the profiles $\phi(\tilde{x})$ and 
$\tilde{\psi}(\tilde{x})$ obtained from a numerical solutions of Eqs. 
(\ref{eq_el1})--(\ref{eq_el3}) for varying surface potentials. The bulk 
composition $\phi_0$ was chosen to reflect a stable vapor phase. 
When $\tilde{V}$ is small ($\tilde{V}=1$ or $2$), the profiles are smooth and 
$\phi(\tilde{x})$ is always close to its bulk value vapor $\phi_0$. In that 
particular example, close to the surface $\phi$ is smaller due to our choice of 
positive value of $\Delta\tilde{\gamma}$. However, if the surface potential is 
large enough ($\tilde{V}=3$, $4$, or $5$), then the electroprewetting phase 
transition pursues marked by a dense liquid phase at the wall and a vapor phase 
far from it. The lower thick dashed line is the linear approximation 
$\phi=\phi_0+\vphi$ with a small $\vphi$ taken from Eq. (\ref{lin_approx}), 
valid before the transition. The higher thick dashed line is a similar 
approximation $\phi=\phi_{b,h}+\vphi$ with values of $\alpha$ and 
$\beta$ from:
\begin{\ea}
\alpha^2&=&\frac{\tilde{f}_{\rm 
vdw}''(\phi_{b,h})-2\tilde{P}\tilde{n}_0(v_0\rho_c\Delta u)^2
e^{v_0\rho_c\Delta u(\phi_{b,h}-\phi_0)}}{\tilde{c}^2},\nn\\
\beta^2&=&\frac{e^{v_0\rho_c\Delta u(\phi_{b,h}-\phi_0)}}
{\tilde{\veps}(\phi_{b,h})}.
\end{\ea}
In Fig. \ref{fig8} we present the total integrated energy of the fluid vs 
$\tilde{V}$ near a slightly hydrophilic surface. The solid line is the 
numerical value of the energy calculated from the sum of Eqs. 
(\ref{dimensionless_bulk_FE}) and (\ref{dimensionless_surf_FE}). The dashed line 
is the analytical approximation Eq. (\ref{quadratic_approx}). The energy 
decreases as  $\tilde{V}$ increases. The numerical and analytical energies are 
quite similar before the critical prewetting potential ($\tilde{V}_c\approx 
4.75$). At $\tilde{V}_c$, the film energy decreases discontinuously and the 
analytical approximation fails.

\begin{figure}[!th]
\begin{center}
\includegraphics[width=0.47\textwidth,bb=60 200 515 580,clip]{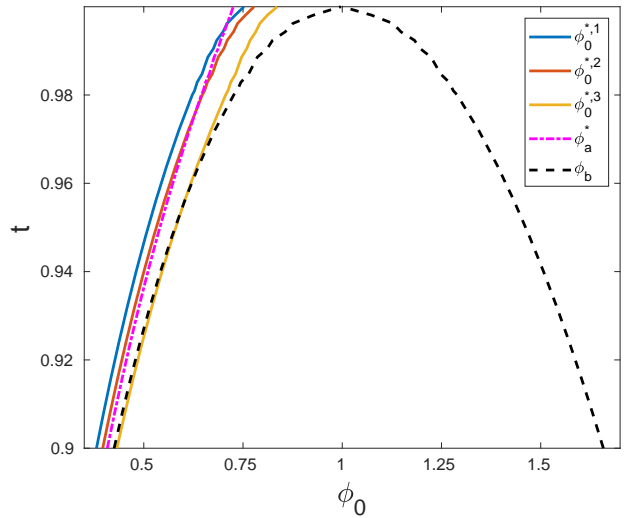}
\caption{Phase diagram in the $(\phi_0,t)$ plane. Dashed line is the classical 
bulk binodal curve $\phi_b$ of the van der Waals model. $\phi_a^*$ is the 
analytical stability line obtained from Eq. (\ref{eq_Vc_sa}) with surface 
potential $\tilde{V}=1$. $\phi_0^{*,1}$ is the phase line for 
$\Delta\tilde{\gamma}=-10^{-3}$ and the same potential. For a given temperature 
and surface potential, $\phi_0^{*,1}$ is the value of $\phi_0$ in Eqs. 
(\ref{eq_el1})--(\ref{eq_el3}) giving $\phi(0)=\phi_{b,l}$. The two curves 
$\phi_0^{*,2}$ and $\phi_0^{*,3}$ are similar but $\Delta\tilde{\gamma}$ is 
$\Delta\tilde{\gamma}=-5\times 10^{-4}$ and $\Delta\tilde{\gamma}=5\times 
10^{-4}$, respectively.
}
\label{fig9}
\end{center}
\end{figure}

Figure \ref{fig9} depicts the phase diagram in the $(\phi_0,t)$ plane with 
five curves. The dashed line is the bulk van der Waals binodal curve. 
$\phi_a^*$ is from the analytical expression from Eq. (\ref{eq_Vc_sa}). 
$\phi_a^*$ is the value of $\phi_0$ satisfying this equation for a given value 
of the surface potential $\tilde{V}=1$ and varying temperatures. Note that due 
to the approximation used, $\phi_a^*$ is not sensitive to the value of 
short-range chemical affinity $\Delta\tilde{\gamma}$. The three other lines are 
the phase lines $\phi_0^{*,i}$ ($i=1$, $2$, $3$) corresponding to the same 
$\tilde{V}$ and three different values of $\Delta\tilde{\gamma}$. They were 
obtained by looking for the value of bulk composition $\phi_0$ that yields a 
numerical solution of Eqs. (\ref{eq_el1})--(\ref{eq_el3}) with 
$\phi_s=\phi_{b,l}$. That is, $\phi_0^{*,i}$ are the values of 
$\phi_0$ where a fluid profile whose density at the surface is exactly equal to 
the (lower) binodal density. When $\Delta\tilde{\gamma}=5\times 10^{-4}$ 
(orange line), the surface is hydrophobic and the unstable region is small. The 
surface hydrophobicity is stronger than the field-effect at all temperature 
except close to the critical point ($t\lesssim 1$). When the surface is 
hydrophilic,  $\Delta\tilde{\gamma}=-5\times 10^{-4}$ (red line), the unstable 
region is larger. Prewetting occurs even farther away from the binodal at 
stronger hydrophilicity, $\Delta\tilde{\gamma}=-10^{-3}$ (blue line).

\section{Conclusions}

We looked at the wetting of a van der Waals fluid model containing dissociated 
ions near a wall with short-range chemical interaction and long-range 
electrostatic forces. Density gradients occur because of two forces: ionic 
screening leads to field gradients and these lead to a dielectrophoretic force 
tending to ``suck'' the fluid to the wall. Simultaneously, ions attracted to the 
wall `drag'' the fluid with them with a force proportional to $\Delta u$ 
stemming from their preferential solvation. We calculated the film energy on the 
mean-field level numerically for all values of surface potential and bulk 
compositions and analytically for conditions where the variations in the density 
profiles are small. 

The phenomenon discussed here has numerous differences from the classical 
wetting. In classical wetting, the first integration of the equation for $\phi$ 
is possible and this gives the profile. The temperature of the ``surface 
critical point'' is different from the bulk critical temperature. In the present 
case, due to the existence of ions and formation of a Debye double layer, the 
integration is not possible and a profile exists even if the gradient-squared 
term is absent from the energy ($\tilde{c}=0$). The long-range electrostatic 
forces give rise to a second-order transition line. A discontinuity in the 
profile requires the inflection of the bulk energy, that is, the first-order 
line terminates at $T_c$. Non-zero preferential solubility ($\Delta u\neq 0$) 
changes this, and so does a nonlinear dependence of $\veps$ on $\phi$, but a 
linear relation $\veps(\phi)$ does not. The second-order line starts from the 
termination of the first-order line and extends towards the bulk critical point. 

The changes to the phase diagram are much larger than those expected by the 
influence of a uniform field (Landau mechanism). The appreciable change in phase 
transition temperature can be seen from the upward shift from the binodal 
temperature, see, for example, Figs. \ref{fig6} and \ref{fig9}. The shift in 
$t$, 
$\Delta t=0.01$--$0.05$, can be translated to a shift in $T$, $\Delta T\simeq 
6$--$30^\circ$K. One can look at the corresponding shift in pressure by 
comparing $\Delta T/T_c$ to $\Delta P/P_c$, and therefore obtain the shift to be 
$\Delta P\approx 10^6$ Pa, that is, about $10$ Atm. The van der Waals model 
employed here is generally considered to give poor quantitative predictions and 
cannot be used to reliably predict the numerical values of many physical 
quantities. Simulations on the molecular level, such as Brownian dynamics or 
Monte Carlo simulations, would allow to obtain more accurate numerical 
predictions. We considered the values of both $\veps$ and $\Delta u$ to be such 
that the liquid is adsorbed at the surface. We leave for a future work to study 
the competition between these forces when their signs are opposite. In the 
Earth's atmosphere, colloidal aggregates and particles are crucial to nucleation 
of liquid droplets. We believe that our results could be implemented to 
ion-induced nucleation around charged particles 
\cite{curtius_nature_2016,kulmala_jgra_2002,froyd_jgra_2004,tsori_jcp_2021,
egorov_prl_2004} and 
possibly shed light on unexplained fast nucleation rates reported.

\begin{acknowledgments}

This work was supported by the Israel Science Foundation Grant No. 274/19.

\end{acknowledgments}

\onecolumngrid

\vspace{1cm}

\begin{center}
\appendix {\bf APPENDIX A}
\end{center}

In this Appendix we expand the free-energy Eq. (\ref{dimensionless_bulk_FE}) to 
quadratic order in $\vphi\equiv \phi-\phi_0$ and $\tilde{\psi}^2$. 
Since $\tilde{n}^\pm\approx \tilde{n}_0(1\mp\psi+v_0\rho_c\Delta u\vphi)$ we 
have
\begin{\ea}
\tilde{n}^+(\ln(\tilde{n}^+)-1)&\approx &
\tilde{n}_0(\ln\tilde{n}_0-1)+\tilde{n}_0\ln\tilde{n}_0v_0
\rho_c\Delta u\vphi+\frac12\tilde{n}_0\left(1+\ln\tilde{n}_0\right)\left(v_0
\rho_c\Delta u\vphi\right)^2\nn\\
&-&\tilde{n}_0\ln\tilde{n}_0\tilde{\psi}-\tilde{n}_0(\ln\tilde{n}_0+1)\tilde{
\psi } v_0\rho_c\Delta u\vphi+\frac12 
\tilde{n}_0(\ln\tilde{n}_0+1)\tilde{\psi}^2,\nn\\
\tilde{n}^-(\ln(\tilde{n}^-)-1)&\approx & 
\tilde{n}_0(\ln\tilde{n}_0-1)+\tilde{n}_0\ln\tilde{n}_0v_0
\rho_c\Delta u\vphi+\frac12\tilde{n}_0\left(1+\ln\tilde{n}_0\right)\left(v_0
\rho_c\Delta u\vphi\right)^2\nn\\
&+&\tilde{n}_0\ln\tilde{n}_0\tilde{\psi}+\tilde{n}_0(\ln\tilde{n}_0+1)\tilde{
\psi } v_0\rho_c\Delta u\vphi+\frac12 
\tilde{n}_0(\ln\tilde{n}_0+1)\tilde{\psi}^2,\nn\\
\Rightarrow\tilde{n}^+(\ln(\tilde{n}^+)-1)&+&\tilde{n}^-(\ln(\tilde{n}^-)-1)\nn
\\=2\tilde{n}_0(\ln\tilde{n}_0-1)&+&2\tilde{n}_0\ln\tilde{n}_0\left(v_0
\rho_c\Delta 
u\vphi\right)+\tilde{n}_0(1+\ln\tilde{n}_0)\left(\tilde{\psi}^2+(v_0
\rho_c\Delta u\vphi)^2\right).
\end{\ea}
\begin{\ea}
(\tilde{n}^+-\tilde{n}^-)\tilde{\psi}\approx 
-2\tilde{n}_0\tilde{\psi}^2-2\tilde{n}_0v_0\rho_c\Delta u\left(\vphi+\frac12 
v_0\rho_c\Delta u\vphi^2\right)\tilde{\psi}^2.
\end{\ea}
\begin{\ea}
\left(\tilde{n}^++\tilde{n}^-\right)\phi&\approx &
\tilde{n}_0\left(1-\tilde{\psi}+v_0\rho_c\Delta 
u\vphi+\frac12(\tilde{\psi}-v_0\rho_c\Delta 
u\vphi)^2\right)(\phi_0+\vphi)\nn\\
&+&\tilde{n}_0\left(1+\tilde{\psi}+v_0\rho_c\Delta 
u\vphi+\frac12(-\tilde{\psi}-v_0\rho_c\Delta 
u\vphi)^2\right)(\phi_0+\vphi)\nn\\
&=&2\tilde{n}_0\phi_0\nn\\
&+&2\tilde{n}_0\vphi+2\tilde{n}_0\phi_0v_0\rho_c\Delta u\vphi\nn\\
&+&\tilde{n}_0\phi_0\left(\tilde{\psi}^2+(v_0\rho_c\Delta 
u\vphi)^2\right)+2\tilde{n}_0v_0\rho_c\Delta u\vphi^2 +O(\vphi^3).
\end{\ea}
Therefore the free-energy Eq. (\ref{dimensionless_bulk_FE}) can be written as
\begin{\ea}
\frac{F}{\lambda_{D0}^3P_c}&\approx 
&\int\left\{\tilde{f}_{\rm vdw}(\phi_0)+ \tilde{f}_{\rm 
vdw}'(\phi_0)\vphi+\frac12 \tilde{f}_{\rm 
vdw}''(\phi_0)\vphi^2
+\frac12\tilde{c}^2(\tilde{\nabla}\vphi)^2 \right.\nn\\
&+&\tilde{P}\left[2\tilde{n}_0(\ln\tilde{n}_0-1)+2\tilde{n}_0\ln\tilde{n}_0
\left(v_0\rho_c\Delta 
u\vphi\right)+\tilde{n}_0(1+\ln\tilde{n}_0)\left(\tilde{\psi}^2+(v_0
\rho_c\Delta u\vphi)^2\right)
\right]\nn\\
&-&\frac12 M\left(\veps(\phi_0)+\Delta 
\veps\vphi\right)(\tilde{\nabla} 
\tilde{\psi})^2
-2\tilde{P}\tilde{n}_0\ln\tilde{n}_0\tilde{\psi}^2-2\tilde{P}\tilde{n}_0v_0
\rho_c\Delta u\vphi\tilde{\psi}^2\nn\\
&-&\left.\tilde{P}v_0
\rho_c\Delta u\left[2
\tilde{n}_0\phi_0+2\tilde{n}_0\vphi+2\tilde{n}_0\phi_0v_0\rho_c\Delta 
u\vphi+\tilde{n}_0\phi_0\left(\tilde{\psi}^2+(v_0\rho_c\Delta 
u\vphi)^2\right)+2\tilde{n}_0v_0\rho_c\Delta u\vphi^2 
\right]
\right\}{\rm d}\tilde{{\bf r}}.\nn
\end{\ea}

To this energy we add the chemical potential terms. They are
\begin{\ea}
\frac{F_\mu}{\lambda_{D0}^3P_c}&=&
\int\left\{-\mu_0(\phi_0+\vphi)-\mu^+\tilde{n}_0\left(1-\tilde{\psi}+v_0\rho_c
\Delta u\vphi+\frac12(\tilde{\psi}-v_0\rho_c\Delta u\vphi)^2\right)\right.\nn\\
&-&\left.\mu^-\tilde{n}_0\left(1+\tilde{\psi}+v_0\rho_c\Delta 
u\vphi+\frac12(\tilde{\psi}+v_0\rho_c\Delta u\vphi)^2\right)\right\}{\rm 
d}\tilde{{\bf r}},\nn
\end{\ea}
where
\begin{\ea}
\mu_0&=&\tilde{f}'_{\rm vdw}(\phi_0)-2\tilde{P}\tilde{n}_0v_0\rho_c\Delta 
u,\nn\\
\mu^\pm&=&\tilde{P}\left(\ln\tilde{n}_0-v_0\rho_c\Delta 
u\phi_0\right).
\end{\ea}
The sum $\Omega=F+F_\mu$ is
\begin{\ea}
\frac{\Omega}{\lambda_{D0}^3P_c}&=&
\int\left\{\tilde{f}_{\rm vdw}(\phi_0)+\tilde{f}_{\rm 
vdw}'(\phi_0)\vphi+\frac12 \tilde{f}_{\rm 
vdw}''(\phi_0)\vphi^2
+\frac12\tilde{c}^2(\tilde{\nabla}\vphi)^2 -\frac12 
M\left(\veps(\phi_0)+\Delta \veps\vphi\right)(\tilde{\nabla} 
\tilde{\psi})^2\right.\nn\\
&-&2\mu^\pm\tilde{n}_0+2\tilde{P}\tilde{n}_0(\ln\tilde{n}_0-1)-\mu_0\phi_0-2
\tilde{P}\tilde{n}_0v_0\rho_c\Delta u\phi_0\nn\\
&+&\left(-\mu_0-2\mu^\pm\tilde{n}_0 v_0\rho_c\Delta u+2\tilde{P}\tilde{n}_0 
v_0\rho_c\Delta u\ln\tilde{n}_0+2\tilde{P}\tilde{n}_0v_0\rho_c\Delta 
u(-v_0\rho_c\Delta u\phi_0-1)\right)\vphi\nn\\
&+&\left(-\mu^\pm\tilde{n}_0 (v_0\rho_c\Delta 
u)^2-\tilde{P}\tilde{n}_0(v_0\rho_c\Delta 
u)^2+\tilde{P}\tilde{n}_0(v_0\rho_c\Delta 
u)^2\ln\tilde{n}_0-\tilde{P}\tilde{n}_0(v_0\rho_c\Delta 
u)^3\phi_0\right)\vphi^2\nn\\
&+&\left[-\tilde{n}_0\left(\mu^\pm+\tilde{P}(1-\ln\tilde{n}_0+v_0\rho_c\Delta 
u\phi_0)\right)-\tilde{n}_0v_0\rho_c\Delta 
u\left(\mu^\pm+\tilde{P}(1-\ln\tilde{n}_0+v_0\rho_c\Delta 
u\phi_0)\right)\vphi\right.\nn\\
&+&\left.\left. \frac12 \tilde{n}_0(v_0\rho_c\Delta 
u)^2\left(\mu^\pm+\tilde{P}(1-\ln\tilde{n}_0+v_0\rho_c\Delta 
u\phi_0)\right)\vphi^2
\right]\tilde{\psi}^2
\right\}{\rm d}\tilde{{\bf r}}.\nn
\end{\ea}
The terms $\tilde{f}_{\rm 
vdw}'(\phi_0)\vphi-\mu_0\vphi-2\tilde{P}\tilde{n}_0v_0\rho_c\Delta u\vphi=0$ 
cancel out, and the surface term $\vphi(0)\Delta\tilde{\gamma}$ is added. The 
result, after additional cleaning, is Eq. (\ref{fe_expansion}).

\vspace{1cm}

\begin{center}
\appendix {\bf APPENDIX B}
\end{center}

In this Appendix we substitute the profiles in Eq. (\ref{lin_approx}) in the 
quadratic free-energy Eq. (\ref{fe_expansion}). We note that 
$\int_0^\infty\tilde{\psi}^2dx=\tilde{V}^2/2\beta$ 
and $\int_0^\infty\tilde{\psi }'^2dx =\tilde{V}^2\beta/2$; 
in addition
\begin{\ea}
\int_0^\infty \vphi^2 
dx&=&\frac{A^2}{2\alpha}-\frac{2AB}{(\alpha-2\beta)(\alpha+2\beta)^2}+\frac{B^2}
{4\beta(\alpha^2-4\beta^2)^2},\nn\\
\int_0^\infty \vphi'^2 dx&=&\frac12 
A^2\alpha-\frac{4AB\alpha\beta}{(2\beta+\alpha)^2(\alpha-2\beta)}+\frac{B^2\beta
}{(4\beta^2-\alpha^2)^2},\nn\\
\int_0^\infty \vphi\tilde{\psi}^2 dx&=&
\frac14 \tilde{V}^2\frac{B+4A\beta(2\beta-\alpha)}{\beta(4\beta^2-\alpha^2)},
\nn\\
\int_0^\infty \vphi\tilde{\psi}'^2 dx&=&
\frac14 
\tilde{V}^2\beta\frac{B+4A\beta(2\beta-\alpha)}{4\beta^2-\alpha^2}.
\nn
\end{\ea}
We write 
$\Omega_0/(\lambda_{D0}^3P_c)=\phi_0\Delta\tilde{\gamma}+\int_0^\infty[
\tilde{f}_{\rm vdw}(\phi_0)-2\tilde{P}\tilde{n}_0-\mu_0\phi_0]{\rm 
d}\tilde{x}$ and use $2\tilde{P}\tilde{n}_0=M$:
\begin{\ea}
\frac{\Omega-\Omega_0}{\lambda_{D0}^3P_c}&=&
\frac12\tilde{c}^2\left[
\frac{A^2\alpha}{2}-\frac{2AB\alpha^2}{(\alpha-2\beta)(\alpha+2\beta)^2}+\frac{B
^2\alpha^2 }
{4\beta(\alpha^2-4\beta^2)^2}\right]\nn\\
&+&\frac12\tilde{c}^2\left[
\frac12 
A^2\alpha-\frac{4AB\alpha\beta}{(2\beta+\alpha)^2(\alpha-2\beta)}+\frac{B^2\beta
}{(4\beta^2-\alpha^2)^2}\right]\nn\\
&-&\frac14 
M\veps(\phi_0)\tilde{V}^2\beta-\frac18 M\Delta 
\veps\tilde{V}^2\beta\frac{B+4A\beta(2\beta-\alpha)}{4\beta^2-\alpha^2}
\nn\\
&-&\frac14M\tilde{V}^2/\beta\nn\\
&-&\frac18Mv_0\rho_c\Delta u
\tilde{V}^2\frac{B+4A\beta(2\beta-\alpha)}{\beta(4\beta^2-\alpha^2)}
\nn\\
&+&\left(A+\frac{B}{4\beta^2-\alpha^2}\right)\Delta\tilde{\gamma}\nn.
\end{\ea}
\begin{\ea}
\frac{\Omega-\Omega_0}{\lambda_{D0}^3P_c}&=&\frac12\tilde{c}^2
A^2\alpha+\frac12\tilde{c}^2\left[
-\frac{2AB\alpha^2+4AB\alpha\beta}{(\alpha-2\beta)(\alpha+2\beta)^2}+\frac{B
^2\left(\alpha^2+4\beta^2\right) }
{4\beta(\alpha^2-4\beta^2)^2}\right]\nn\\
&-&\frac18 M\Delta \veps
\tilde{V}^2\beta\frac{B+4A\beta(2\beta-\alpha)}{4\beta^2-\alpha^2}\nn\\
&-&\frac12M\tilde{V}^2/\beta\nn\\
&-&\frac18Mv_0\rho_c\Delta u
\tilde{V}^2\frac{B+4A\beta(2\beta-\alpha)}{\beta(4\beta^2-\alpha^2)}
\nn\\
&+&\left(A+\frac{B}{4\beta^2-\alpha^2}\right)\Delta\tilde{\gamma}\nn.
\end{\ea}
Further simplification yields:
\begin{\ea}
\frac{\Omega-\Omega_0}{\lambda_{D0}^3P_c}&=&\frac12\tilde{c}^2\left(
A^2\alpha-\frac{2AB\alpha}{\alpha^2-4\beta^2}+\frac{B^2\left(\alpha^2+4\beta^2
\right)}{4\beta\left(\alpha^2-4\beta^2\right)^2}\right)\nn\\
&-&\frac18 M 
\tilde{V}^2\left(\Delta\veps\beta^2+v_0\rho_c\Delta 
u\right)\frac{B+4A\beta(2\beta-\alpha ) } 
{\beta \left(4\beta ^2- \alpha^2\right) }\nn\\
&-&\frac12M\tilde{V}^2/\beta\nn\\
&+&\left(A+\frac{B}{4\beta^2-\alpha^2}\right)\Delta\tilde{\gamma}\nn.
\end{\ea}
We now use the definition of $B=-M\left(v_0\rho_c\Delta u+\Delta\veps\beta^2 
\right)\tilde{V}^2/2\tilde{c}^2$ on the second line:
\begin{\ea}
\frac{\Omega-\Omega_0}{\lambda_{D0}^3P_c}&=&\frac12\tilde{c}^2\left(
A^2\alpha-\frac{2AB\alpha}{\alpha^2-4\beta^2}+\frac{B^2\left(\alpha^2+4\beta^2
\right)}{4\beta\left(\alpha^2-4\beta^2\right)^2}\right)\nn\\
&+&\frac14 B\tilde{c}^2\frac{B+4A\beta(2\beta-\alpha ) } 
{\beta \left(4\beta ^2- \alpha^2\right) }\nn\\
&-&\frac12M\tilde{V}^2/\beta\nn\\
&+&\left(A+\frac{B}{4\beta^2-\alpha^2}\right)\Delta\tilde{\gamma}\nn.
\end{\ea}
We combine the first and second lines:
\begin{\ea}
\frac{\Omega-\Omega_0}{\lambda_{D0}^3P_c}=\frac12\tilde{c}^2\left(
A^2\alpha-\frac{4AB\beta}{\alpha^2-4\beta^2}-\frac14\frac{B^2\left(
\alpha^2-12\beta^2\right)}{\beta\left(\alpha^2-4\beta^2\right)^2}
\right)
-\frac12M\tilde{V}^2/\beta
+\left(A+\frac{B}{4\beta^2-\alpha^2}\right)\Delta\tilde{\gamma}\nn.
\end{\ea}
Equation (\ref{en_integ}) is retrieved.

\twocolumngrid

\bibliography{references.bib}

\end{document}